\documentclass[structabstract]{aa}
\usepackage{txfonts}
\usepackage{epsfig}
\usepackage{multirow}
\usepackage[section] {placeins}
\usepackage{graphicx}

\begin{document}
\title{Automated period detection from variable stars'\\ time series database}
\author{Shaju.K.Y. \inst{1}
\thanks{\emph{email:}
shajuky@gmail.com, sky@cusat.ac.in}
\and Piet Reegen \inst{2}
\and Ramesh Babu Thayyullathil \inst{1}}
\institute{Cochin University of Science and Technology, Cochin, Kerala, India
\and Institute of Astronomy, University of Vienna, Austria }

\abstract
{The exact period determination of a multi-periodic variable star based on its luminosity time series data is believed a task requiring skill and experience. Thus the majority of available time series analysis techniques require human intervention to some extent.}
{The present work is dedicated to establish an automated method of period (or frequency) determination from the time series database of variable stars.}
{Relying on the \sc SigSpec \rm method (Reegen 2007), the technique established here employs a statistically unbiased treatment of frequency-domain noise and avoids spurious (i.\,e. noise induced) and alias peaks to the highest possible extent. Several add-ons were incorporated to tailor \sc SigSpec \rm to our requirements. We present tests on 386 stars taken from ASAS2 project database.}
{From the output file produced by \sc SigSpec, \rm the frequency with maximum spectral significance is chosen as the genuine frequency. Out of 386 variable stars available in the ASAS2 database, our results contain 243 periods recovered exactly and also 88 half periods, 42 different periods etc.}
{\sc SigSpec \rm has the potential to be effectively used for fully automated period detection from variable stars' time series database. The exact detection of periods helps us to identify the type of variability and classify the variable stars, which provides a crucial information on the physical processes effective in stellar atmospheres.}
\maketitle


\section{Introduction}

Iterative measurements of a quantity over time yield a time series. This work deals with time series obtained from photometric observations of variable stars, which is the photometric flux (or magnitude) versus time, with or without data gaps. The majority of astronomical measurements cannot be taken continuously over long periods of time due to several reasons. For ground based observations, daylight and the weather conditions are unavoidable sources of data gaps, and observations from space may suffer from cosmic particle impacts or stray light corruption occasionally producing data points beyond repair (Reegen et al. 2006).

The method to extract physical information from time-resolved data is commonly known as time series analysis. The present paper is primarily dedicated to regular variables, which exhibit a strictly periodic photometric signal, so the primary goal of our analysis will be to identify the appropriate frequency in case of a mono-periodic star. If the detected frequency is exact, the intrinsic scatter of a phased light curve will attain a minimum. For multi-periodic variables, a step-by-step prewhitening procedure has established and is widely used in combination with multi-sine least-square fitting techniques (Sperl 1998; Lenz \& Breger 2005). The identified periods permit to identify the type of variability and deduce important astrophysical parameters. The Fourth Variable Star Working Group meeting held at Geneva Observatory, Switzerland in 2005 elaborates about the Period Search Benchmarks (Laurent Eyer)\footnote{\tt http://obswww.unige.ch/\textasciitilde eyer/VSWG/Meeting4/MINUTES/g\\aia-vswg4.html}.

\section{Various astronomical surveys}

Modern observational astronomy yields a huge amount of observational time series data available from various survey observations such as ASAS \footnote{\tt http://www.astrouw.edu.pl/asas/ } (All Sky Automated Survey -- detection of photometric variability), MACHO \footnote{\tt http://wwwmacho.anu.edu.au/  } (MAssive Compact Halo Objects -- search for dark matter by gravitational lensing), EROS \footnote{\tt http://eros.in2p3.fr/ } (Exp\'erience pour la Recherche d'Objets Sombres -- search for and study of dark stellar bodies by their gravitational microlensing effects on stars), OGLE \footnote{\tt http://ogle.astrouw.edu.pl/ }  (Optical Gravitational Lensing Experiment -- search for dark matter by gravitational microlensing), ROTSE \footnote{\tt http://www.rotse.net/ }(Robotic Optical Transient Search Experiment -- searching gamma-ray bursts),  The PLANET   \footnote{\tt http://bustard.phys.nd.edu/MPS/ }Collaboration (Probing Lensing Anomalies NETwork -- detecting and characterising microlensing anomalies), MISAO  \footnote{\tt http://www.aerith.net/misao/ } Project (Multitudinous Image-based Sky-survey and Accumulative Observations -- making use of images in the world for new object discoveries and data acquisition of known objects), Pan-STARRS  \footnote{\tt http://pan-starrs.ifa.hawaii.edu/public/ }(Panoramic Survey Telescope And Rapid Response System - asteroids, comets, variable stars), LSST  \footnote{\tt  http://www.lsst.org/lsst }(Large Synoptic Survey Telescope - variable sources, Transient alerts) and in the near future, more data are expected from the ESA \footnote{\tt http://sci.esa.int/science-e/www/area/ind\\ex.cfm?fareaid=1 } mission Gaia (three-dimensional map of the Milky Way), APASS  \footnote{\tt  http://www.aavso.org/apass}(AAVSO Photometric All-Sky Survey) etc. Even though these data are collected on different purposes, the data analysis of these time series can lead to the discovery of interesting new objects and variable stars.

\section[]{Existing methods and their limitations}

Various approaches existing at present to extract frequencies from a variable star's time series data, most of which are listed at the Geneva University website\footnote{\tt http://obswww.unige.ch/\textasciitilde eyer/VSWG/tools.html}. Many of them are based on the Discrete Fourier Transform (DFT), which can be regarded as a correlation between the measured time series and trigonometric functions -- sines and cosines, with frequency as the independent parameter. The consideration of cosine and sine to represent a two-dimensional Fourier vector defines the Fourier space, and the normalisation of the cosine and sine covariances provides the length of the Fourier vector to return the signal amplitude. A plot of amplitude versus frequency is termed as the amplitude spectrum of the time series. Peaks in an amplitude spectrum indicate frequencies where the dataset correlates better with the trigonometric functions than elsewhere, and the idea of period detection is to assume that the highest peaks indicate signals produced by the star, whereas the lower peaks are due to random measurement errors and are frequently called noise. The simplest method to distinguish between signal and noise is to average the amplitudes over a certain frequency range and to compare the peak amplitude to this environmental mean (Breger et al. 1993). One of the potential drawbacks of employing such a signal-to-noise ratio is that it depends on the frequency range used for averaging and -- more critically -- on (yet) unresolved signal components hidden in the noise.

The problem of finding the appropriate noise level in the amplitude spectrum -- or the degree of randomness in a time series -- was addressed by several authors providing different solutions. Some introduce corrections to the DFT itself, such as the Lomb-Scargle Periodogram (Lomb 1976; Scargle 1982) or the Date Compensated DFT (Ferraz-Mello 1981), others apply statistical methods such as ANOVA analysis (Schwarzenberg-Czerny 1996, 1997). As an alternative to trigonometric functions, Akerlof et al. (1994) examined cubic spline fits for time series analysis.

A completely different way of period detection is the systematic examination of phased light curves modulo different periods and to determine the best-fitting period by minimising the intrinsic scatter of the phase plot. This was introduced as the string-length method by Lafler \& Kinman (1965). The most common formal representation is the Phase Dispersion Minimisation (Stellingwerf 1978). This method does not require any initial assumption on the shape of the periodicity and works also for non-trigonometric signals, but its application to multiperiodic data suffers from the systematic influence of unresolved signal components on the statistical behavior of the phase plots. François Mignard's FAMOUS\footnote{\tt ftp://ftp.obs-nice.fr/pub/mignard/Famous/} is a recent method, which uses a sinusoidal model to fit the data with the amplitude coefficients being either constant or a polynomial in time.

Most of the methods described above are effective in finding ``true'' frequencies, but mixed with spurious or alias frequencies with almost comparable amplitudes. There are some useful software packages like {\sc PERIOD04}\footnote{\tt http://www.univie.ac.at/tops/Period04/}, {\sc PERANSO}\footnote{\tt http://tonnyvanmunster.ipage.com/peranso/downloads.htm} and {\sc MUFRAN}\footnote{\tt www.konkoly.hu/tifran}, but they require human supervision and intervention in between, so that automation is not easy. Although these programs are well-suitable for analysing a single star's data, they become extremely time-consuming if applied to huge time series databases. Hence in the era of petabytes of survey data, it is price-worthy to fully automate the process of frequency detection.

\begin{figure}
\centering
\begin{tabular}{c}
\epsfig{file=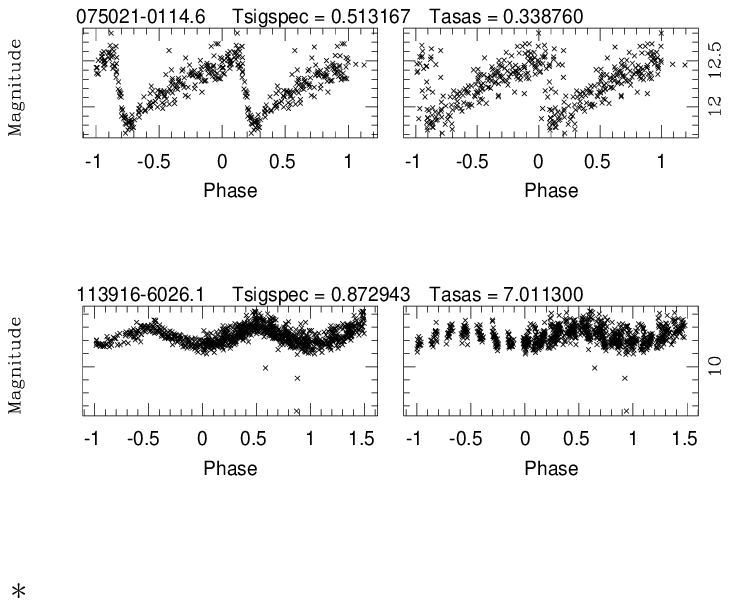,width=0.8\linewidth,height=.8\linewidth,bb=75 580 312 746 ,clip=true}
\end{tabular}
\caption{Periods and light curves of two stars, where the {\sc SigSpec} period differs from the previously published value. The {\sc SigSpec} (left) light curve is better than that from ASAS (right). See Table\,\ref{TABdiff}}\label{FIGdiff}
\end{figure}

\section[]{Frequency Analysis by SigSpec}

The principle of {\sc SigSpec} (Significance Spectrum) is described by Reegen (2007). The ANSI-C source code of {\sc SigSpec} as well as the user manual (Reegen--2010) can be downloaded at the {\sc SigSpec} website\footnote{\tt http://www.sigspec.org}. The linux script and new add-on subroutines are available from the author. The method is based on the analytical solution for the frequency-domain Probability Density Function (PDF) of white noise, which depends on amplitude, frequency and phase, and on least-squares fitting of trigonometric functions to data. It has several advantages compared to the currently used Lomb-Scargle Periodogram, Phase Dispersion Minimization (PDM) and other Discrete Fourier Transform (DFT) methods. One of the advantages is that it defines a mathematical quantity called ``Spectral Significance`` ({\it sig}) and thereby another quantity, known as ``Cumulative Significance``({\it csig})  which permits to identify the most reliable period without any statistical bias. Through an iterative procedure, {\sc SigSpec} detects the most significant frequency component and performs the corresponding prewhitening to find the next dominant frequency. We have modified {\sc SigSpec} to handle global settings in addition to individual ones referring to each individual dataset. This additional option enables the software to handle multiple files in a fully automatic mode. For the present task, the {\sc SigSpec} runs were encapsulated in a Linux shell script, and the appropriate handling of each light curve was realized through global adjustments. Thus {\sc SigSpec} can now be used for an automated frequency search, since the dominant frequency is selected merely by a quantitative comparison rather than human visual inspection, which may be qualitative and biased in nature.

The initial attempts of the most of the time series data analysts is, through the visual examination on the time series data through respective plots to obtain the Nyquist sampling frequency and some other parameters needed for further analysis. Since we know that any kind of human intervention is time-consuming for huge databases and is a barrier for automation, our attempt was to avoid such manual interruptions at any stage.

\section[]{Application to ASAS Data}

To show the accuracy and efficiency of {\sc SigSpec}, under automation, we chose the ASAS2 project database\footnote{\tt http://www.astrouw.edu.pl/asas}, which contains time series data files of 386 variable stars. The same database was employed for an automated frequency analysis previously by Min-Su et al. (2004). The ANSI-C source code and the scripts needed to run the Min-Su's program {\sc MS\_Period} can also be downloaded at his website \footnote{\tt http://www.astro.lsa.umich.edu/\textasciitilde msshin/scien\\ce/code/MultiStep\_Period}. Even though that was the previous published work aiming automation of period detection, the method does not define a mathematical or statistical quantity, based on which the true period can be selected. The quantities used in that approach, such as AoV or $\chi^2$-statistic will be of less help for choosing the single true period quantitatively by a computer program. For each star, the method gives the possible ten candidate periods and their light curves. Afterwards the true period has to be confirmed again by visual inspection of the light curves. However Min-Su's paper explicitly describes the need for automation and the related uncertainties involved during the automation process.

\begin{figure}
\centering
\begin{tabular}{c}
\epsfig{file=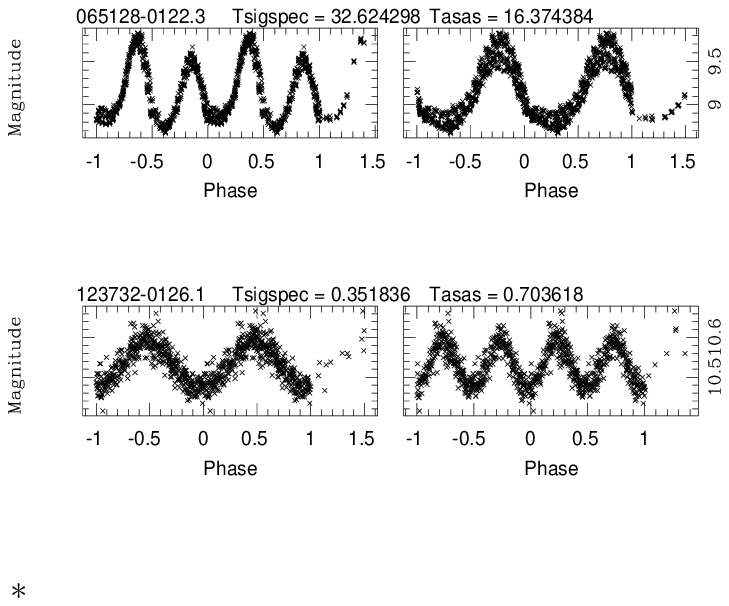,width=0.8\linewidth,height=.8\linewidth,bb=75 580 312 746 ,clip=true}
\end{tabular}
\caption{Periods and light curves of two stars, where the {\sc SigSpec} period is twice the previously published value. The SigSpec (left) light curve is better than that from ASAS (right).  See Table\,\ref{TABmulti}}\label{FIGdoub}
\end{figure}

\subsection[]{{\sc SigSpec} setup}

Our proposed procedure starts with the application of {\sc SigSpec} to all 386 time series files, one by one. The automatic creation of proper runtime environment for running the {\sc SigSpec} over each time series file is taken care by a Linux script, which also creates the required ''SigSpec.ini'' file for each time series. The {\sc SigSpec} default settings are applied, except to the following four parameters, which has to be specified in the ``SigSpec.ini`` file during the run-time. They are
\begin{enumerate}
\item {\tt lfreq} - The lower frequency limit was set to $\frac{1}{T}$, where $T$ is the total time span of the time series.
\item {\tt nycoef} - In the chosen ASAS database, the time-domain sampling of several datasets is sparse compared to the expected frequency according to Min-su et al.~(2004). To overcome the ambiguity in defining a Nyquist Frequency for non-equidistant sampling, Reegen (2010) introduces the Nyquist Coefficient. Each sampling interval of width $\delta t_k = t_k - t_{k-1}$ is considered to have its individual Nyquist frequency $\left(2\delta t_k\right)^{-1}$. For an upper frequency limit $f_\mathrm{u}$, the Nyquist Coefficient is defined as the fraction of sampling intervals for which $\left(2\delta t_k\right)^{-1} > f_\mathrm{u}$. By default, the program {\sc SigSpec} assigns the value of the Nyquist Coefficient as $0.5$, which may be reasonable in many situations and to restrict the frequency values in the expected range. For the specified ASAS database, the Nyquist Coefficient was given as $0.9$ in order to set a larger value for the upper frequency {\tt ufreq}. The total run-time for each time series and hence for the entire database depends mainly on this value.
\item {\tt iterations} - By default, {\sc SigSpec} stops its calculations, if the {\it sig} drops below 5. For our tasks, we used a fixed number of 5 iterations to keep the computation time reasonable\footnote{For the 386 targets under consideration, the entire period finding process took 3 hours and 48 minutes on an Intel Core 2 CPU 2.8GHz Linux PC} and because our task was to compare the dominant frequencies to the values from the ASAS database.
\item {\tt siglimit} - In some cases, before completing the 5 iterations specified above, the {\it sig} value may drops below 5, and the program may not give the required 5 results. In order to avoid this situation, the default significance limit of 5 was modified. Thus in order to detect weaker signal components in the case of noisier datasets, a {\it siglimit} of 2 was specified.
\end{enumerate}

Apart from these, two major add-ons to {\sc SigSpec} were developed to improve both the performance and our results.
\begin{itemize}
\item The five frequencies with the highest cumulative sig are taken into account to identify the fundamental in a possible set of harmonics. In some of the cases, it was found that the cumulative sig for the fundamental frequency is lower than that of an integer multiple. In order to compensate this, we introduced an add-on which picks the fundamental frequency and brings it as the first, by checking all the integral overtones among each other.
\item For all ground-based observations, $1\,$d$^{-1}$ alias will be present and hence the statistical behavior of the sampling characteristics in Fourier Space is taken into account appropriately by {\sc SigSpec}. However, ground-based single-site data additionally suffer from nightly zero-point uncertainties. These invoke signals in the domain of $1\,$d$^{-1}$ and integer multiples, which are intrinsic to the time series. Consequently, {\sc SigSpec} considers these pseudo-signals significant in the sense of not due to noise. To overcome this potential problem in our automated analysis, we explicitly put a condition to avoid frequencies between $0.99\,$d$^{-1}$ and $1.01\,$d$^{-1}$ while selecting the frequency with maximum sig from the results of {\sc SigSpec}.
\end{itemize}

\section{Results and Discussion}

Among the 386 ASAS datasets under consideration, there was one corrupt file (135546-2911.5) and one time series file consisting of only two measurements (052927-6852.0). These two files were rejected initially, and we performed the fully automated frequency analysis for the remaining 384 stars. The summary of our results are as follows.\\

\begin{itemize}
\renewcommand{\labelitemi}{$\bullet$}
\item Correct Periods   :  243
\item Half Periods        :    88
\item Quarter Periods   :     4
\item Double Periods    :     3
\item Different Periods :   42
\item Not Published      :     4
\item Corrupt files         :     2
\end{itemize}

For 243 stars (\textasciitilde64\%), our results are exactly identical to the published frequencies.

For 95 stars, (88+4+3 = 95) (\textasciitilde25 \%) (Table\,\ref{TABmulti}), the {\sc SigSpec} frequencies are integer or half-integer multiples of the published frequencies. Out of these:
\begin{itemize}
\item in 88 cases, the {\sc SigSpec} frequency is twice the published frequency,
\item in 4 cases, the {\sc SigSpec} frequency is 4 times the published frequency,
\item in 3 cases, the {\sc SigSpec} frequency is half the published frequency,
\end{itemize}

This is not necessarily to be interpreted as a disagreement, but may rather indicate the presence of a non-sinusoidal waveform, which is represented by a fundamental frequency plus integer harmonics in Fourier space. In these 95 cases, {\sc SigSpec} assigns a higher statistical significance to a different representative of the same non-sinusoidal periodicity as published.

\begin{figure}
\flushleft
\begin{tabular}{c}
\epsfig{file=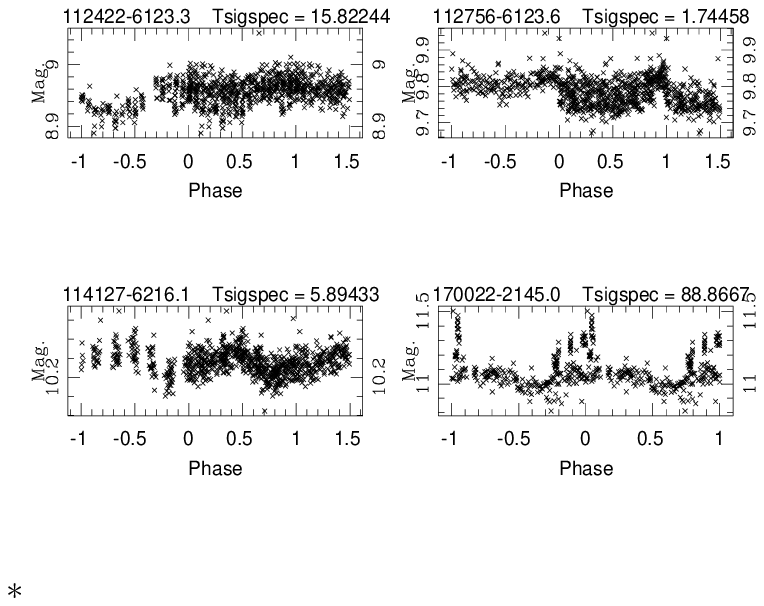,width=0.86\linewidth,height=.8\linewidth,bb=75 580 312 746 ,clip=true}
\end{tabular}
\caption{ Periods and Light curves of four stars, whose periods are not published by ASAS. See Table\,\ref{TABnew}}\label{FIGnew}
\end{figure}

The 42 periods (\textasciitilde11\%) flagged as ``different'', no reasonable assignment to integer or half-integer multiples of the published values is possible. Among these datasets, 27 are strongly contaminated with alias frequencies of $1\,$d$^{-1}$. Table\,\ref{TABalias} shows that the {\sc SigSpec} results represent various linear combinations of the published frequency and $1\,$d$^{-1}$, such as $f+1$, $f-1$, $2f+1$, $2f-1$, etc., where $f$ is the published frequency. Since the ASAS data represent ground based measurements, it is not surprising that we obtained $1\,$d$^{-1}$ alias frequencies of the published results, which represent an ambiguity common to all time series incorporating periodic gaps. This drawback can only be overcome by continuous measurements over a sufficiently long period of time and is rather intrinsic to the data than a particular weakness of the applied method, as addressed earlier by Reegen (2007). The analysis in terms of frequency instead of period reveals such combinations of true frequencies with the alias frequency. For the remaining 15 stars, there is no obvious agreement between the {\sc SigSpec} result and the published frequency (Table\,\ref{TABdiff}). Out of these 15 results, two phased light curves are displayed in Figure\,\ref{FIGdiff}, which shows that the periods and light curves obtained by the {\sc SigSpec} method are better than the published ones. 

For 4 stars, no frequency was published in the ASAS catalogue and the {\sc SigSpec} frequencies are given in  (Table\,\ref{TABnew}). The corresponding light curves are displayed in Figure\,\ref{FIGnew}. As indicated in Tab.\,\ref{TABnew}, the time interval covered by the measurements is sufficient for a reasonable phase coverage, and the sig values are fairly high. Thus the periodicity is very probably present in the data. The first three stars with star-ID's 112422-6123.3, 114127-6216.1 and 112756-6123.6 are Cepheid variables with periods ranging from 1 to 16 days and amplitude variations from 0.1 to 0.2 magnitudes. The fourth light curve corresponding to the star-ID,  170022-2145.0 shows that, it is a Recurrent Novae, which is similar to Novae, but with few small outbursts during the observed time span. 

\section{Conclusions}
We investigated several existing programs and software packages which are used for the time series analysis of variable stars and found that human intervention is required in some way to confirm the true period. Full automation could be achieved only with {\sc SigSpec} combined with few additional subroutines. The majority of results are found to be satisfying with the published frequencies. Thus {\sc SigSpec} turned out to be a useful utility for future massive variability surveys and also for the re-analysis of already existing time series data of periodic variables.

\section*{Acknowledgments}
{We are grateful for the useful discussions with Ninan Sajeeth Philip and Min-Su Shin. Our thanks are due to Ajit Kembavi, IUCAA, Pune and V.C.Kuriakose, IRC, CUSAT, Kochi, for the facilities provided.  This work was supported by F.I.P. fellowship of U.G.C.,New Delhi, Govt. of India.}


\newpage
\begin{table*}
 \flushleft
 \begin{minipage}{120mm}
\caption{Published frequencies (ASAS Catalog) compared with {\sc SigSpec} frequencies for 95 variable stars incorporating integer frequency multiples.  See Figure\,\ref{FIGdoub}}\label{TABmulti}
 \begin{tiny}
\begin{tabular}{|l|r|r|r|l|r|r|r|}
  \hline
  \hline
\multirow{2}{*}{}Star$-$ID & Published ($f$) & {\sc SigSpec} & Remark \vline & Star$-$ID & Published ($f$) & {\sc SigSpec} & Remark \\
& d$^{-1}$ &d$^{-1}$ & \vline & & d$^{-1}$ & d$^{-1}$ & \\
  \hline
  \hline

065128$-$0122.3 & 0.061071 & 0.030652  &  $0.5f$ \vline    &   174611$+$0043.8 & 0.038806 & 0.019437   &   $0.5f$  \\
065311$+$0033.3 & 0.005874 & 0.002940  &  $0.5f$ \vline    &                  &          &                  &      \\
\hline
\hline
005759$+$0034.7 & 0.626543 & 1.253096   &   $2f$  \vline    &   114416$-$6142.9 & 0.271176 & 0.542458   &   $2f$  \\  
030201$-$0027.2 & 0.160958 & 0.321955   &   $2f$  \vline    &    114417$-$5942.7 & 0.197128 & 0.394082   &   $2f$  \\
045017$+$0100.7 & 2.432049 & 4.864181   &   $2f$  \vline    &    114557$-$6352.9 & 1.048535 & 2.097050   &   $2f$  \\
045206$-$7043.9 & 0.213253 & 0.426546   &   $2f$  \vline    &    114617$-$6100.1 & 0.814267 & 1.628439   &   $2f$  \\
045728$-$7033.1 & 0.606081 & 1.212173   &   $2f$  \vline    &    114720$-$6155.1 & 0.799412 & 1.600792   &   $2f$  \\
045817$-$0013.9 & 1.481306 & 2.960463   &   $2f$  \vline    &    114757$-$6225.3 & 2.529065 & 5.058099   &   $2f$  \\
050047$-$7029.8 & 2.581584 & 5.163185   &   $2f$  \vline    &    114806$-$6221.3 & 0.242099 & 0.484223   &   $2f$  \\
053936$-$7958.6 & 0.543101 & 1.086050   &   $2f$  \vline    &    115934$-$8546.0 & 1.635946 & 3.271839   &   $2f$  \\
055850$-$0026.5 & 0.883939 & 1.886515   &   $2f$  \vline    &    123732$-$0126.1 & 1.421226 & 2.842237   &   $2f$  \\
064914$-$0036.1 & 0.956494 & 1.912960   &   $2f$  \vline    &    123748$-$6219.4 & 2.299369 & 4.598663   &   $2f$  \\
064926$-$0013.6 & 0.455409 & 0.910743   &   $2f$  \vline    &    123808$-$6353.8 & 0.888200 & 1.776485   &   $2f$  \\
065633$-$0104.9 & 0.914411 & 2.008088   &   $2f$  \vline    &    123824$-$6404.8 & 0.130913 & 0.259711   &   $2f$  \\
075623$-$0043.7 & 0.868835 & 1.737659   &   $2f$  \vline    &    124145$-$6241.4 & 0.046908 & 0.093989   &   $2f$  \\
095706$-$0120.7 & 1.013125 & 2.026236   &   $2f$  \vline    &    124221$-$6259.6 & 0.530617 & 1.061186   &   $2f$  \\
103617$-$5202.5 & 2.368495 & 4.737194   &   $2f$  \vline    &    124351$-$6305.2 & 0.079898 & 0.160235   &   $2f$  \\
103843$-$5245.9 & 0.534499 & 1.069123   &   $2f$  \vline    &    124435$-$6331.7 & 0.389842 & 0.779472   &   $2f$  \\
104447$-$5155.1 & 0.141926 & 0.283586   &   $2f$  \vline    &    125124$-$6404.7 & 0.159190 & 0.317747   &   $2f$  \\
104526$-$5224.3 & 0.419526 & 0.839116   &   $2f$  \vline    &    125210$-$6312.7 & 0.126117 & 0.246145   &   $2f$  \\
112229$-$6313.7 & 0.256331 & 0.512619   &   $2f$  \vline    &    125319$-$6401.4 & 2.347688 & 4.695407   &   $2f$  \\
112330$-$6423.3 & 1.104704 & 2.209445   &   $2f$  \vline    &    125816$-$6258.1 & 0.396082 & 0.792166   &   $2f$  \\
112419$-$6344.0 & 0.031052 & 0.062196   &   $2f$  \vline    &    125827$-$6230.9 & 0.164338 & 0.328795   &   $2f$  \\
112421$-$5917.1 & 0.290988 & 0.582139   &   $2f$  \vline    &    125933$-$6210.5 & 0.647874 & 1.295804   &   $2f$  \\
112550$-$5935.6 & 0.260869 & 0.521635   &   $2f$  \vline    &    125953$-$6159.5 & 0.674466 & 1.333530   &   $2f$  \\
112645$-$6251.8 & 0.626944 & 1.253954   &   $2f$  \vline    &    131314$-$8528.5 & 1.808953 & 3.617803   &   $2f$  \\
112742$-$6127.8 & 1.114859 & 2.229744   &   $2f$  \vline    &    135340$-$3036.0 & 2.101485 & 4.202853   &   $2f$  \\
112746$-$6110.5 & 0.621876 & 1.239722   &   $2f$  \vline    &    135734$-$3139.1 & 0.202082 & 0.404312   &   $2f$  \\
112826$-$5929.4 & 0.222272 & 0.444369   &   $2f$  \vline    &    144245$-$0039.9 & 0.465007 & 0.930099   &   $2f$  \\
112852$-$6255.8 & 0.198062 & 0.396122   &   $2f$  \vline    &    163954$-$0033.2 & 0.867032 & 1.733911   &   $2f$  \\
112901$-$6052.8 & 0.615032 & 1.230073   &   $2f$  \vline    &    174135$-$0035.7 & 0.235014 & 0.470343   &   $2f$  \\
112924$-$6154.5 & 0.622158 & 1.244267   &   $2f$  \vline    &    174619$-$0018.7 & 0.714466 & 1.428966   &   $2f$  \\
112927$-$6201.9 & 0.310116 & 0.620240   &   $2f$  \vline    &    175502$-$2314.3 & 0.382462 & 0.764478   &   $2f$  \\
112939$-$5953.7 & 0.521805 & 1.043548   &   $2f$  \vline    &    175859$-$2323.1 & 0.649131 & 1.298251   &   $2f$  \\
113210$-$5948.9 & 2.205120 & 4.410282   &   $2f$  \vline    &    180253$-$2409.6 & 0.474136 & 0.948294   &   $2f$  \\
113252$-$6228.1 & 0.579568 & 1.159037   &   $2f$  \vline    &    184139$-$0044.7 & 3.477983 & 5.954052   &   $2f$  \\
113321$-$5949.6 & 0.882708 & 1.765316   &   $2f$  \vline    &    185448$-$2326.2 & 0.283689 & 0.567574   &   $2f$  \\
113451$-$6128.0 & 0.292426 & 0.584923   &   $2f$  \vline    &    185621$-$4040.8 & 0.669414 & 1.338900   &   $2f$  \\
113648$-$6425.6 & 0.472214 & 0.944369   &   $2f$  \vline    &    190403$-$2228.6 & 0.737400 & 1.474814   &   $2f$  \\
113713$-$5952.6 & 0.732555 & 1.464975   &   $2f$  \vline    &    195723$-$2105.2 & 2.181644 & 4.363257   &   $2f$  \\
113746$-$6014.6 & 0.339191 & 0.678498   &   $2f$  \vline    &    204045$+$0056.4 & 0.422274 & 0.844557   &   $2f$  \\
114035$-$6306.2 & 3.497152 & 6.994285   &   $2f$  \vline    &    204859$+$0027.4 & 1.947469 & 3.894899   &   $2f$  \\
114226$-$6228.6 & 0.683397 & 1.366813   &   $2f$  \vline    &    210554$-$1647.8 & 3.296120 & 5.589594   &   $2f$  \\
114257$-$6248.4 & 1.120834 & 2.241778   &   $2f$  \vline    &    220248$-$1218.7 & 3.259739 & 6.519366   &   $2f$  \\
114304$-$5956.4 & 0.225743 & 0.451451   &   $2f$  \vline    &    225935$-$0702.4 & 0.062448 & 0.124831   &   $2f$  \\
114346$-$6144.6 & 0.334125 & 0.668200   &   $2f$  \vline    &    & &  & \\
\hline
\hline
045128$-$0032.7 & 0.276327 & 1.279096   &   $4f$  \vline    &     113333$-$6353.7 & 1.008980 & 4.035765   &   $4f$  \\
112356$-$6105.4 & 0.218327 & 0.945571   &   $4f$  \vline    &     113612$-$6317.0 & 0.166987 & 0.667287   &   $4f$  \\
\hline
\hline
\end{tabular}
\end{tiny}
\end{minipage}
\end{table*}

\begin{table*}
 \flushleft
 \begin{minipage}{110mm}
\caption{Published frequencies (ASAS Catalog) are not available for comparison with {\sc SigSpec} frequencies for 4 variable stars.  See Figure\,\ref{FIGnew}} \label{TABnew}
\begin{tabular}{|l|r|r|r|l|r|r|r|}
  \hline
  \hline
\multirow{2}{*}{}Star$-$ID & {\sc SigSpec} & time span & sig \vline & Star$-$ID  & {\sc SigSpec} & time span & sig \\
&  d$^{-1}$ & d & \vline & &  d$^{-1}$ & d &  \\
\hline
\hline
112422-6123.3& 0.063201&1013.9&27.44   \vline & 114127-6216.1 & 0.169646&1015.0&32.07  \\
112756-6123.6 & 0.573202&1015.0&22.03   \vline & 170022-2145.0 & 0.011252&909.8&20.60  \\
\hline
\hline
\end{tabular}
\end{minipage}
\end{table*}

\begin{table*}
 \flushleft
 \begin{minipage}{120mm}
\caption{Published frequencies (ASAS Catalog) compared with {\sc SigSpec} frequencies for 15 variable stars with different results. See Figure\,\ref{FIGdiff}} \label{TABdiff}
\begin{tabular}{|l|r|r|l|r|r|}
  \hline
  \hline
\multirow{2}{*}{}Star$-$ID & Published & {\sc SigSpec} \vline &Star$-$ID   & Published & {\sc SigSpec}  \\
&  d$^{-1}$ & d$^{-1}$ \vline & & d$^{-1}$ &  d$^{-1}$ \\
\hline
\hline

050556$-$6810.7 & 0.569152 & 0.067644    \vline &  052832$-$6836.2 & 2.234872 & 1.733874\\
065608$-$0059.1 & 0.204876 & 2.073359    \vline &  075554$-$0016.8 & 0.551937 & 0.003426\\     
104755$-$5214.9 & 0.565804 & 0.008316    \vline &  112325$-$6238.7 & 0.107230 & 0.003234\\     
112815$-$5932.4 & 0.005424 & 1.025093    \vline &  112923$-$6241.4 & 0.768167 & 0.003547\\     
113426$-$6320.0 & 0.582330 & 4.005417    \vline &  113457$-$6157.6 & 0.005026 & 0.160237\\     
114248$-$5859.6 & 0.082109 & 0.065919    \vline &  114308$-$6029.1 & 0.645444 & 0.070683\\     
114726$-$6132.9 & 0.012755 & 0.194331    \vline &  125922$-$6217.3 & 0.459303 & 0.058550\\     
204430$-$0028.6 & 2.346630 & 0.168650    \vline  &  & & \\
\hline
\hline
\end{tabular}
\end{minipage}
\end{table*}

\begin{table*}
 \flushleft
 \begin{minipage}{120mm}
\caption{Published frequencies (ASAS Catalog) compared with {\sc SigSpec} frequencies for 27 variable stars incorporating $1\,$d$^{-1}$  alias. }\label{TABalias}
   \begin{tabular}{|l|r|r|r|l|r|r|r|}
  \hline
  \hline
\multirow{2}{*}{}Star$-$ID & Published & {\sc SigSpec} & Remark \vline & Star$-$ID & Published & {\sc SigSpec} & Remark \\
& d$^{-1}$ & d$^{-1}$ & \vline &  &  d$^{-1}$ &  d$^{-1}$ & \\
  \hline
  \hline
055602$-$0003.8  &  0.450672  &  0.101238   &    1$-$2f  \vline & 112706$-$6037.3  &  0.047339  &  0.141980   &    3f \\
055624$+$0013.0  &  0.126196  &  0.757343   &    1$-$2f  \vline & 113917$-$6210.5  &  0.079164  &  0.237435   &    3f \\
064858$-$0037.6  &  0.370872  &  0.260917   &    1$-$2f  \vline & 114250$-$6226.1  &  0.158169  &  0.474540   &    3f \\
112301$-$6146.8  &  0.373633  &  0.253018   &    1$-$2f  \vline & 015647$-$0021.2  &  2.845379  &  1.842615   &    f$-$1 \\
150439$-$1536.3  &  0.491407  &  0.018985   &    1$-$2f  \vline & 075021$-$0114.6  &  2.951942  &  1.948682   &    f$-$1 \\
104838$-$5245.7  &  0.864125  &  0.140362   &    1$-$f  \vline & 134460$-$3019.2  &  1.140287  &  0.139034   &    f$-$1 \\
114303$-$6150.6  &  0.019280  &  0.983377   &    1$-$f  \vline & 184343$-$0014.2  &  2.808571  &  1.806973   &    f$-$1 \\
114658$-$6045.4  &  0.033314  &  0.966597   &    1$-$f  \vline & 200208$-$1958.8  &  1.044968  &  0.045283   &    f$-$1 \\
180206$-$3554.5  &  0.018052  &  0.983247   &    1$-$f  \vline & 060013$+$0046.9  &  0.091285  &  1.091264   &    f$+$1 \\
112720$-$5955.8  &  0.023457  &  0.006750   &    2f$-$1  \vline & 065210$-$0017.8  &  0.470593  &  1.477265   &   f$+$1 \\
112803$-$6411.4  &  0.835197  &  0.664193   &    2f$-$1  \vline & 113916$-$6026.1  &  0.142627  &  1.145550   &    f$+$1 \\
064851$+$0022.7  &  0.090627  &  1.181365   &    2f$+$1  \vline & 123639$-$6344.8  &  0.665336  &  1.665406   &    f$+$1 \\
064918$-$0003.5  &  1.292673  &  3.588224   &    2f$+$1  \vline & 185355$-$4038.4  &  0.026976  &  1.028787   &    f$+$1 \\
213919$-$0106.5  &  0.157129  &  1.316912   &    2f$+$1  \vline &  & & & \\
\hline
\hline
\end{tabular}
\end{minipage}
\end{table*}

\end{document}